# Novel development of dissipative-soliton-resonance pulses with pump power in an all-normal-dispersion fiber laser

Yufei Wang,[1] Lei Li,[1] Shuai Wang,[1] Liming Hua,[1] Chaojie Shu,[1] Lei Su,[2] D. Y. Tang,[1] D. Y. Shen,[1] Luming Zhao,[1,2,*]

1. Jiangsu Key Laboratory of Advanced Laser Materials and Devices, Jiangsu Collaborative Innovation Center of Advanced Laser Technology and Emerging Industry, School of Physics and Electronic Engineering, Jiangsu Normal University, Xuzhou, Jiangsu 221116 China
2. School of Engineering and Materials Science, Queen Mary University of London, London, UK
*Corresponding author: lmzhao@ieee.org

**Evolution of dissipative-soliton-resonance (DSR) pulses in an all-normal-dispersion fiber laser with pump power under various operation conditions are experimentally studied. The fiber laser is mode-locked by using a nonlinear fiber loop mirror. Apart from the typical pulse broadening due to the peak power clamping effect, pulse breaking was observed. In addition, pulse narrowing with pump power increasing was observed. All the incompatible evolution of DSR pulses could be attributed to the multiple parameters change due to pump power increasing under specific operation conditions.**

http://dx.doi.org/

## 1. INTRODUCTION

Passively mode-locked fiber lasers have been widely reckoned as effective tools for generations of ultrashort pulse due to their compact, versatile, alignment-free features. Particularly, the pursuit of high-energy pulse output from passively mode-locked lasers has never ceased in both academic and industrial fields. However, wave breaking restricts the pulse energy boost because of the excessive accumulated nonlinear effect, leading to the pulse splitting. Multiple pulses have been observed and discussed in different passively mode-locked fiber lasers [1–5].

To circumvent wave breaking operation, dissipative soliton resonance (DSR), a square-like pulse formation mechanism theoretically proposed by Chang et al in 2008 [6], has attracted intense attentions in the past decade. Within the DSR regime, the peak of the pulse remains constant while the pulse width broadens linearly with increasing pump power. The DSR phenomenon indicates that the pulse energy can be indefinitely high as long as the pulse width is arbitrarily broadened without pulse splitting. Both theoretical analysis and a variety of experiments have proved that under a proper setup of cavity parameters, the generation of DSR is independent of cavity dispersion and mode-locking techniques [7–19]. Multipulse operations of DSR harmonic mode-locking are also investigated theoretically [20] and experimentally [21–24]. Komarov et al claimed that the number of DSR pulses in steady state is related to the initial conditions, regardless of the increasing pump power [20]. In practice, the pump power increasing may not solely increase the effective gain. Many parameter changes are resulted from pump power increasing, such as central wavelength moving, cavity birefringence change, CW modification etc. Therefore, DSR pulse evolution with pump power may exhibit various performance, even opposite to the theoretical prediction.

In this paper, we report on the pulse dynamics of single DSR pulse evolution with pump power in an all-normal dispersion fiber laser with a nonlinear fiber loop mirror (NOLM). At first, the DSR feature of broadening pulse duration with constant pulse peak power is observed. Further increasing the pump power, we find that the DSR performance will not be maintained while keeping other laser configuration unchanged. The DSR pulse will evolve into several specific states such as dual-pulse and harmonic mode-locking. The pulse width will even narrow as the pump power increases. The mode-locked state will eventually be lost due to the infinite increase of pump power. It looks like a procedure of DSR splitting. However, carefully analysis suggest that it is a new DSR generation rather than the pulse splitting of the original DSR. To the best of our knowledge, this is the first demonstration of the unsustainability of DSR pulse broadening due to the impact of pump power on the laser parameters.

## 2. EXPERIMENTAL SETUP

The all-normal-dispersion (ANDi) fiber laser proposed in our experiment is schematically shown in Fig. 1. A NOLM, which acts as a fast saturation absorber, is coupled into a unidirectional ring (UR) cavity through a 30/70 fiber coupler (OC). The NOLM consists of a polarization controller (PC1) and a piece of 200-m-long single-mode fiber (SMF, Nufern, 1060-XP) to enhance the nonlinear effect. In the UR, the gain is provided by a 30-cm-long single-cladding, ytterbium-doped fiber (YDF, Coractive Yb501) with core absorption 139 dB/m at 915 nm. The YDF is pumped through a wavelength division multiplexer (WDM) by a 976 nm pigtailed laser diode, providing a maximum pump power of 600 mW. A polarization-independent isolator (PI-ISO) is inserted into the UR to ensure the unidirectional operation and prevent backwards reflections. Another polarization controller (PC2) is used to adjust the state of intra-cavity polarization. And a fiber coupler with 20% port acting as an output is placed after PC2. The fiber pigtails of all the optical components in this laser are HI1060. The total length of the laser is 229 m, corresponding to a fundamental repetition rate of ~900 KHz. A 63 GHz high speed oscilloscope (Agilent, DSA-X 96204Q) together with a 45-GHz photodetector (New Focus, 1014) is employed to visualize the output pulse train. The spectrum is analyzed by an optical spectrum analyzer (OSA, Yokogawa AQ6317C). The output power is monitored by an optical power meter (Agilent, 81681A). The radio-frequency (RF) spectra are measured by a RF-spectrum analyzer (Agilent, N9320B).

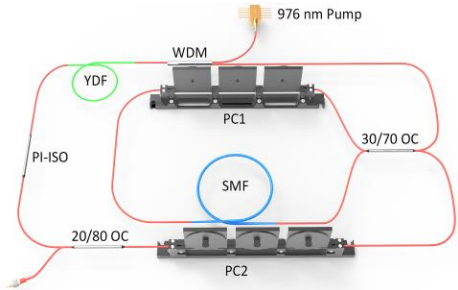

**Fig. 1.** Schematic of the all-normal-dispersion fiber laser. WDM, wavelength division multiplexer; YDF, ytterbium-doped fiber; PI-ISO, polarization-independent isolator; PC, polarization controller; OC, fiber coupler; SMF, 200-m-long single mode fiber.

## 3. EXPERIMENTAL RESULTS AND DISCUSSION

In our experiment, by appropriately adjusting the orientation of the polarization controllers, mode-locked square pulse operating in DSR regime can always be easily obtained at a pump threshold of about 40 mW. The mode-locked pulse can be sustained due to the pump hysteresis effect [3,25,26] when we decrease the pump power to 24.5 mW. The oscilloscope trace of the mode-locked pulse under the pump power of 24.5 mW is shown in Fig. 2(a), depicting that the measured pulse duration is about 896 ps. The inset in Fig. 2(a) shows that the pulse train has a uniform interval of 1.11 μs, corresponding to the cavity roundtrip time. The single-pulse temporal profile exhibits an asymmetrical structure as the pulse front has higher intensity than the pulse trailing edge. Meanwhile, Fig. 2(b) shows a bell-shaped optical spectrum of the mode-locked pulse. The result corresponds to previous reports [21,27–30], and can be deduced that the central part of spectrum shares more gain than the edges. The laser emission is centered at 1034.84 nm with a 0.079 nm 3-dB spectral bandwidth. The calculated time bandwidth product (TBP) is 19.8, manifesting that the pulse is heavily chirped. According to the radio frequency (RF) spectrum illustrated in Fig. 2(c), the signal-to-noise ratio exceeds 60 dB at 900 kHz. The inset in Fig.2(c) with a broader span of 100 MHz confirms that stable mode-locking operation is achieved.

With fixed PC paddles, the square pulse broadens gradually when we increase the pump power from 24.5 mW to 37.7 mW as can be seen in Fig. 3(a). The increase of the pulse duration starts at the pulse trailing edge which makes the top of the pulse evolve into an extended plateau. The pulse front does not change and still exhibits higher intensity. Fig. 3(b) is the evolution of the optical spectrum. It is notable that the central wavelength shifts from 1034.84 nm to 1036.47 nm, and the 3-dB spectral bandwidth slightly changes from 0.079 nm to 0.047 nm. A slight increase of spectral intensity can be observed. Fig. 3(c) presents the average output power, pulse widths, pulse energy and peak power as a function of the pump power. The pulse duration increases monotonically from 896 ps to 2.187 ns versus the pump power. Correspondingly, the average output power and pulse energy also increase linearly with the pump power, while the peak power of the pulse remains almost constant in the process of broadening. Note that there is no fine structure or unstable states observed in this phase of evolution, suggesting that the pulse operates in the DSR regime.

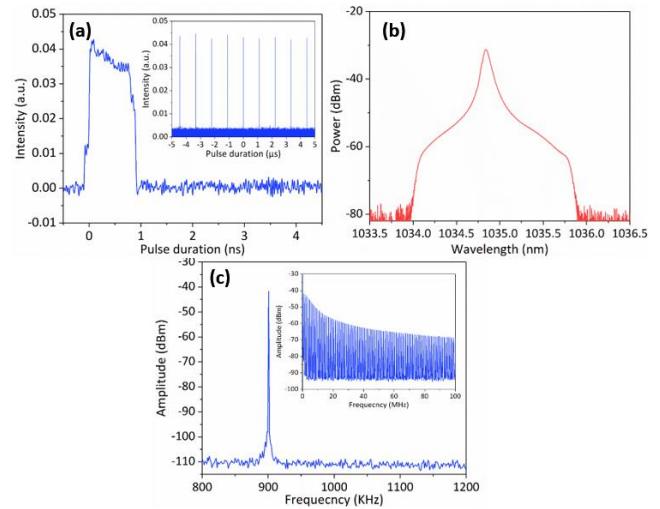

**Fig. 2.** Typical square pulse emission under 24.5 mW pump power. (a) Oscilloscope trace of a single pulse. Inset: the pulse train. (b) Corresponding optical spectrum. (c) RF spectrum with a span of 400 kHz and a resolution bandwidth of 10 Hz. Inset: 100 MHz span, 1KHz resolution bandwidth

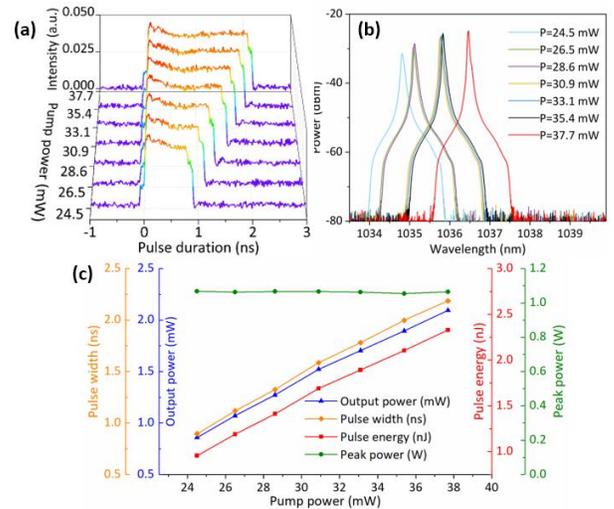

**Fig. 3.** Increasing the pump power from 24.5 mW to 37.7 mW (a) Dynamics of pulse broadening operates in DSR regime. (b) Corresponding optical spectrum with increasing pump power. (c) Pulse width, output pump power, pulse energy and peak power versus the pump power.

By continuously increasing the pump power from 39.8 mW to 76.3 mW, the evolution of the pulse temporal and spectral profile is further investigated. Surprisingly, although the pulse keeps broadening as the pump power increases, its DSR characteristics can no longer be maintained. In the time domain, as shown in Fig. 4(a), the pulse duration is unable to maintain linear growth. The intensity of the flat plateau slightly increases accompany with the pump power while the pulse front is gradually weakened which introduces changes to the asymmetrical structure of the original pulse. The pulse width increases from 2.438 ns to 3.533ns. Vivid changes also emerged in the optical spectrum within this pump range, as shown in Fig. 4(b). Under the pump power of 39.8 mW, a secondary signal centered at ~1085 nm appears in the optical spectrum apart from the main signal centered at 1036.49 nm, which is resulted from the stimulated Raman scattering (SRS) effect [16, 31–33]. The long cavity configuration may contribute to the low threshold for the onset of SRS [34]. The increase of the pump power not only accelerates the rise of the SRS components, but also causes the blue-shift of both signal components. Previous report about long cavity laser configuration in ANDi [31] suggests that the advent of SRS can affect the energy scalability as it shares part of the pump energy. And the aggravated SRS can destabilize the pulse in the form of sporadic bursts of intense fluctuations. Moreover, when the pump power reaches 67.1 mW, a new peak located around 1039 nm rises and surely steals the energy of main signal. The changes in the optical spectrum with pump power boost may explain the evolution of temporal pulse profile. The curve of the output power and pulse width as a function of pump power shows obvious nonlinearity as can be seen in Fig. 4(c). The variation of peak power indicates the loss of DSR characteristic clearly. Both the oscilloscope trace and optical spectrum show that the pulse evolves into an unstable state when the pump is increased to 76.3 mW. The broadening process is terminated and the average output power changes dramatically.

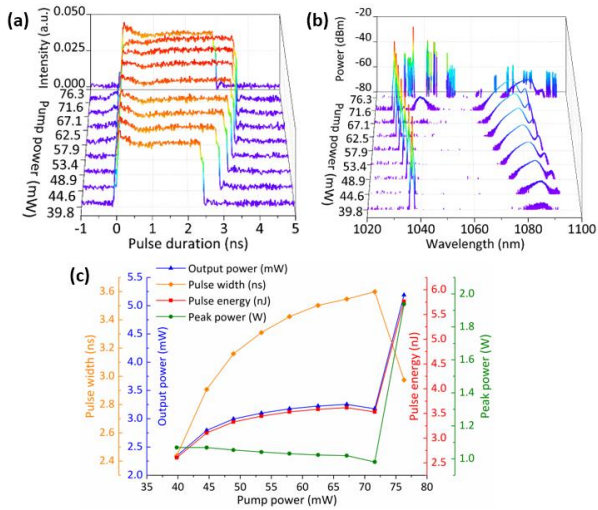

**Fig. 4.** Increasing the pump power from 39.8 mW to 76.3 mW (a) Oscilloscope trace of pulse evolution. (b) Corresponding optical spectrum. (c) Pulse width, output pump power, pulse energy and peak power versus the pump power.

Slightly increasing the pump power to 78.5 mW, as shown in Fig. 5(a), the unstable oscilloscope trace automatically transforms into a steady dual-pulse state. The dual-pulse duration is about 5 ns, the narrower sub-pulse duration is about 0.55 ns and the wider sub-pulse duration is about 3.55 ns. The sub-pulse separation is about 0.9 ns. The inset in Fig. 5(a) shows that the pulse pair has an interval of 1.11 μs, in agreement with Fig. 2(a). Fig. 5(b) shows the corresponding RF spectrum. The repetition rate of the pulse pair is still 900 kHz. Adding the pump power from 78.5 mW to 90.1 mW, the evolution of the dual-pulse can be seen in Fig. 5(c) and Fig. 5(d). In the time domain, the structure of the dual-pulse does not change significantly, except when the pump power is increased to 85.4 mW, the pulse duration of narrower sub-pulse becomes about 0.9 ns. The spectral component located around 1039 nm gradually rises from the CW substrate as the pump power increases. But the blue shift and ridge of the spectral component are discontinuous, a jump occurs when the power reaches 85.4 mW.

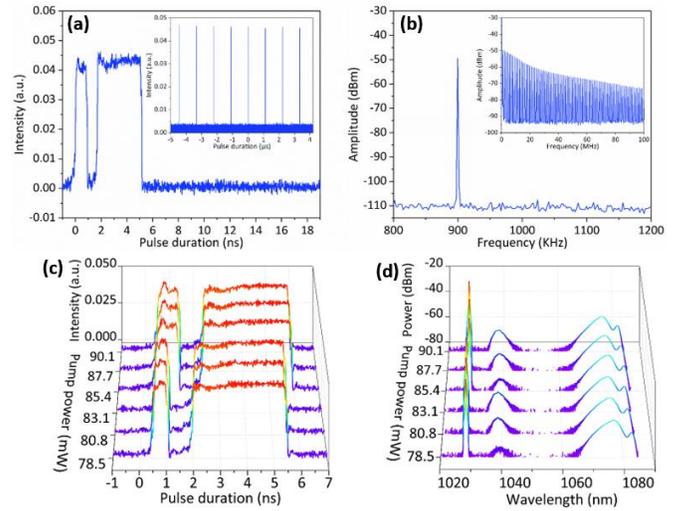

**Fig. 5.** Increasing the pump power from 78.5 mW to 90.1 mW (a) Oscilloscope trace of a dual-pulse. Inset: the pulse train under 78.5 mW. (b)RF spectrum with a span of 400 kHz and a resolution bandwidth of 10 Hz. Inset: 100 MHz span, 1KHz resolution bandwidth (c) Oscilloscope trace of pulse evolution. (d) Corresponding optical spectrum.

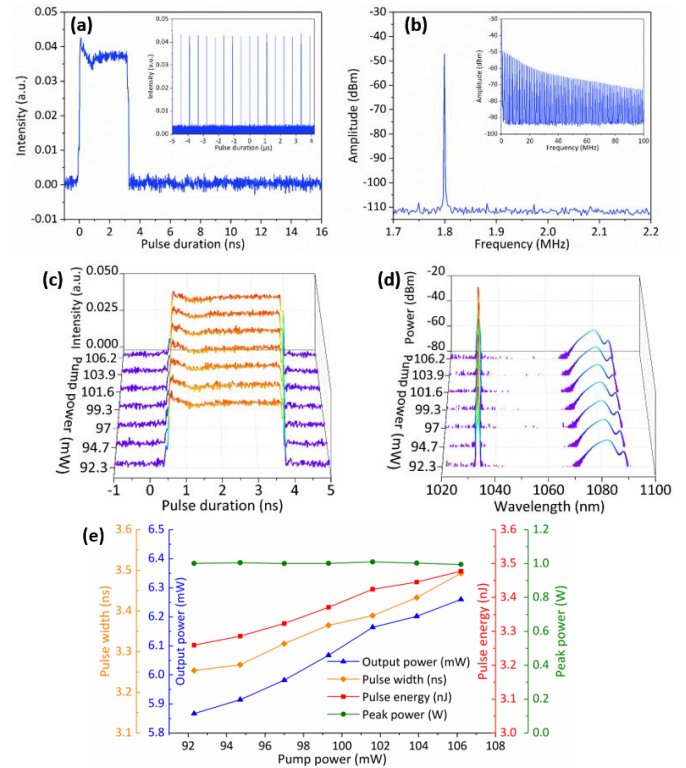

**Fig. 6.** Increasing the pump power from 92.3 mW to 106.2 mW (a) Oscilloscope trace of a harmonic mode-locking pulse. Inset: the pulse train under 92.3 mW. (b)RF spectrum with a span of 400 kHz and a resolution bandwidth of 10 Hz. Inset: 100 MHz span, 1KHz resolution bandwidth (c) Oscilloscope trace of pulse evolution. (d) Corresponding optical spectrum. (e) Pulse width, output pump power, pulse energy and peak power versus the pump power.

Continue to increase the pump power to 92.3 mW, the dual-pulse state evolves into a 2-order harmonic mode locking (HML) of single square pulse as shown in Fig. 6(a). As presented in the inset of Fig. 6(a), the interval of uniform pulse train becomes 555.6 ns which is half of the original DSR pulse interval. In Fig. 6(b), measured RF spectrum shows a signal to noise ratio around 60 dB at 1.8 MHz. In Fig.6(c), when we increase the pump power from 92.3 mW to 106.2 mW, the square pulse broadens without wave-breaking during the process. Compared to the spectrum in the dual-pulse regime, evident changes are reflected in Fig. 6(d) as the pump power increases. The center wavelength of the pulse is switched from 1027.3 nm to 1033.3 nm during the transition of the pulse from dual-pulse state to HML. The CW ridge around 1039 nm disappears while the rise of SRS around 1080 nm still exists. Fig.6(e) indicates that the measured pulse duration still increases quasi-linearly with the pump power and the peak power remains almost constant.

When the pump power exceeds 106.2 mW, the HML pulse will be unstable again. Unrestricted pump power boost will eventually result in the loss of mode-locked state. To ensure that this DSR pulse evolution as pump power increases is not accidental, we start to reduce pump power from 106.2 mW, and the above phenomenon will still appear in order. The pump power is increased and decreased multiple times, and the evolution of the DSR pulse discussed above is consistent.

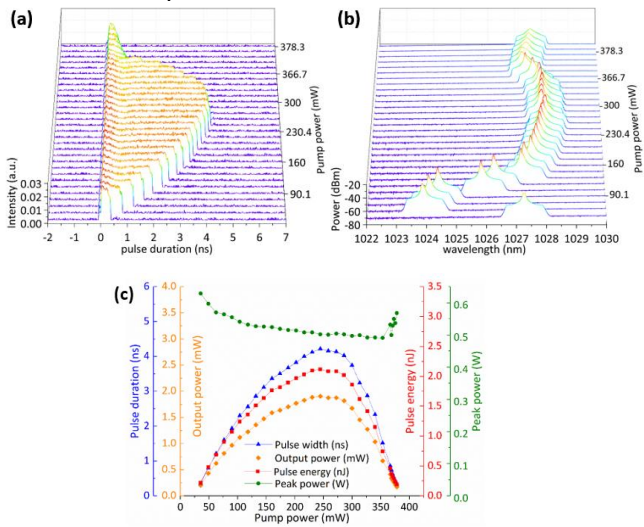

**Fig. 7.** Tuning pump power from 35.64 mW to 244.5 mW (a) Dynamics of pulse broadening and shrinking operates in DSR regime. (b) Corresponding optical spectrum with increasing pump power. (c) Pulse width, output pump power, pulse energy and peak power versus the pump power.

In order to further investigate the influence of pump power on the DSR pulses, we also change the rotation of the PCs to study the pulse evolution path under different intra-cavity birefringence conditions. Under a specific PC configuration in our laser, pulse duration of the DSR pulse also cannot increase monotonically with pump power boost. Fig.7(a) depicts a special evolution of temporal profile of DSR pulse versus increasing pump power. It can be clearly seen from the figure that the broadening of the DSR pulse cannot maintain linearity as the pump power increases and the saturation trend of pulse duration occurs when the pulse is broadened to a certain extent. The peak of the pulse remains constant during the entire broadening process. From 35.64 mW to 244.5 mW, the pulse duration changes from 355 ps to 4.214 ns. When the power exceeds 244.5 mW, It is noticeable that the pulse no longer continues to broaden but begins to shrink. As the pump power increases, the pulse contraction is faster than the broadening process. The change of spectrum is also depicted in the Fig.7(b). The entire process is accompanied by significant wavelength-shift. Unlike the pulse evolution previously discussed, no new frequency components appear on the spectrum. The long NOLM configuration introduces a sinusoidal transfer functions which acts like a saturable absorber and causes the strong peak-power-clamping effects [35]. We can deduce that the anti-saturable absorption is not fully suppressed as the pump power increases and leads to the inverse modulation of the pulse shape. More details about pulse duration, average power, peak power and pulse energy are depicts in Fig.7(c). The pulse broadening process will gradually saturate, and continuing to increase the pump power will cause the pulse width to shrink rapidly. The peak power fluctuations appearing at the beginning and end of the whole process are caused by the asymmetric structure of the pulse.

## 4. CONCLUSION

In conclusion, we have experimentally demonstrated novel evolution of DSR pulses with pump power in an Yb-doped NOLM-based all-normal-dispersion fiber laser. We found that the increase of pump power will affect multiple intra-cavity parameters of the laser. Simply increasing the pump power cannot guarantee that the DSR pulse can be infinitely and linearly broadened. Under appropriate operation condition, the original DSR pulse may evolve into a new DSR pulse with different center wavelength, multi-pulses, or HML pulses. The change of the birefringence in the cavity may even cause shrinkage of the DSR pulse with pump power increasing. This greatly enriches our understanding of the dynamics of DSR pulses.